\documentclass[%
 onecolumn,
superscriptaddress,
 amsmath,amssymb,
 aps,
longbibliography
]{revtex4-2}

\usepackage{graphicx}
\usepackage{dcolumn}
\usepackage{bm}
\usepackage{hyperref}
\usepackage[mathlines]{lineno}

\usepackage{soul}
\usepackage{float}

\newcommand{\comment}[1]{}

\usepackage[usenames, dvipsnames]{color}

\begin{document}

\preprint{APS/123-QED}

\title{Nonreciprocal model swimmer at intermediate Reynolds numbers}


\author{Hong T. Nguyen} 
\affiliation{Department of Applied Physical Sciences, The University of North Carolina at Chapel Hill, Chapel Hill, North Carolina 27599, USA}

\author{Daphne Klotsa} 
\email{dklotsa@email.unc.edu}
\affiliation{Department of Applied Physical Sciences, The University of North Carolina at Chapel Hill, Chapel Hill, North Carolina 27599, USA}

\begin{abstract}
Metachronal swimming, the sequential beating of limbs with a small phase lag, is observed in many organisms at various scales, but has been studied mostly in the limits of high or low Reynolds numbers. 
Motivated by the swimming of brine shrimp, a mesoscale organism that operates at intermediate Reynolds numbers, we computationally studied a simple nonreciprocal 2D model that performs metachronal swimming. Our swimmer is composed of two pairs of paddles beating with a phase difference that are symmetrically attached to the sides of a flat body. We numerically solved the Navier-Stokes equations and used the immersed boundary method to model the interactions between the fluid and swimmer. To investigate the effect of inertia and geometry, we performed simulations varying the paddle spacing and the Reynolds numbers in the range $Re = 0.05 - 100$. In all cases, we observed back-and-forth motion during the cycle and a finite cycle-averaged swim speed at steady state. Interestingly, we found that the swim speed of the swimmer has nonmonotonic dependence on $Re$, with a maximum around $Re\approx1$, a flat minimum between $Re=20-30$ and an eventual increase for $Re>35$. To get more insight into the mechanism behind this relationship, we first decomposed the swim stroke of the swimmer and each paddle into power and recovery strokes and characterized the forward and backward motion. Unlike for reciprocal swimmers, here, for parts of the cycle there is competition between the two strokes of the paddles - the balance of which leads to net motion. We then studied the cycle-averaged, as well as, power-stroke-averaged and recovery-stroke-averaged fluid flows, and related differences in the fluid field to the nonmonotonic behavior of the swim speed. Our results suggest the existence of distinct motility mechanisms that develop as inertia increases within the intermediate-Re range.

\end{abstract}

\pacs{Valid PACS appear here}
\maketitle

\section{Introduction}
The locomotion of aquatic organisms is heavily controlled by hydrodyanmics, which varies greatly as the organisms grow in size and undergo physiological transitions, such as growth of limbs. At early hatching stages, small swimmer larvae operate at low Reynolds numbers (Re), where the effect of viscosity dominates and inertia can be neglected (Stokes regime, Re$\ll$1). In contrast, at later development stages, and larger organisms operate at high Re, where the effect of inertia dominates and inviscid theory can be applied (Eulerian regime, Re$\gg1$). Situated between the two limits is the intermediate-Re regime, where both viscous and inertial effects play a role, $Re_{int}\approx 1-500$. Therefore, from hatching to maturity, many aquatic animals have to transition through, or some may spend their entire life cycle in, the intermediate-Re region. Moreover, at intermediate Re, we see a plethora of motility mechanisms (e.g. rowing, flapping, anguilliform, jet-propulsion), and organisms that even transition between different mechanisms as they grow. For example, the mollusk \textit{C.~antartica} switches from using cilia to flapping as it grows ~\cite{childress2004transition}, the brine shrimp transitions from rowing to gliding with metachronally-beating legs~\cite{Williams1994}, and the nymphal mayfly transitions from rowing to flapping with its gill plates~\cite{sensenig2009rowing}. 
Despite this wealth of behavior, our understanding of the hydrodynamics governing motility in the $Re_{int}$ region is limited and has mostly focused on specific organisms~\cite{Bartol2009,Herschlag2011,Kern2006,fuiman1988ontogeny,McHenry2003,Strickler1975,Blake1986,Borrell2005,gemmell2015tale,jiang2011does,wilhelmus2014observations,Nawroth2014,jones2016bristles}. From a physics point of view, it is important to develop minimal models in order to identify and understand general, unifying principles and physical mechanisms at intermediate scales~\cite{Klotsa2019Perspective}.

A model organism for studying motility at $Re_{int}$ is \textit{Artemia}, also known as brine shrimp, because it naturally operates at those scales.  Brine shrimp are part of the crustacean
family, live in high salinity lakes and have no natural
predators. A small, round nauplius ($<1$mm in size) hatches from the
egg, that “lurches” around with just a pair of antennae acting like
oars. As the animal grows, a series of limbs develop
sequentially reaching 11 on either side. 
With age, the swimming speed increases and the
quality of the swimming transitions from “jerky” to
“smooth”~\cite{Williams1994AModel,Anufriieva2014TheSwimmingBehavior}. Adults (size $\approx 1$cm) swim using their limbs
coordinated in a metachronal beating, \textit{i.e.} beating with a phase difference~\cite{Williams1994AModel}. 

Metachronal
waves have been studied in cilia at low Re~\cite{Elgeti2013,brumley2012,brumley2015,Hayashi2020Metachronal} and in crustaceans
at low~\cite{Takagi2015Swimming,Hayashi2020Metachronal} and high Re~\cite{Alben2010Coordination,Murphy2011Metachronal} but less so in the intermediate range. Brine shrimp are a natural choice and experiments have been performed to understand their motility in both adults and larvae~\cite{Williams1994AModel, Williams1994Locomotion, Gauld1959Swimming, Anufriieva2014TheSwimmingBehavior}. Williams~\cite{Williams1994Locomotion} quantified the beating frequency, swimming kinematics and ontogenesis development of brine shrimp during the sequential addition of active limbs along their trunk in the nauplius and metanauplius stages. Anufriieva \textit{et. al.}~\cite{Anufriieva2014TheSwimmingBehavior} supported and extended Williams' findings to later stages of brine shrimp adulthood both in natural and laboratory settings. Recent work characterized the flocking behavior of a population of brine shrimp in response to external stimuli~\cite{Ali2011Complexity, Gulbrandsen2001ArtemiaSwarming}.

Mathematical modeling of metachronal swimming in brine shrimp, and in  general crustaceans, can be classified into two main approaches: mechanical and hydrodynamic~\cite{Byron2021MetachronalMotion}. The mechanical ignores explicit hydrodynamic interactions between body, limbs and surrounding fluid by solving Newton's equations of motion with empirical input parameters from experiments. Williams~\cite{Williams1994AModel} showed that a mechanical model of brine shrimp with just a single beating pair of limbs using an empirical relation of drag coefficients is sufficient to explain the hydrodynamics of larval brine shrimp swimming and that the unsteady forces (i.e. inertial ) became important at later stages when many pairs of limbs beat metachronally. Similarly, using a general drag coefficient model, where fluid forces on the body are assumed to be proportional to the velocity, Alben et. al.~\cite{Alben2010Coordination} showed the advantage of metachronal over synchronous appendage coordination for krill, and in general. However, many parameters need to be accurately estimated from experiments and used as input in the mechanical model, making its applicability somewhat limited. The drawbacks are resolved in the second approach. The hydrodynamic model explicitly includes the hydrodynamic coupling between body, limbs and fluid, with which one can vary and study a host of parameters: body geometry (number of appendages, their length or spacing between them), kinematic variables (beating frequency, phase lag, stroke amplitude), viscosity of fluid, body and fluid inertia characterized by Reynolds numbers. To include explicit hydrodynamics at intermediate Re means that in general the Navier-Stokes (NS) equations need to be solved numerically. Zhang \textit{et. al.}~\cite{Zhang2014Neural, Granzier-nakajima2020ANumerical} simulated a 2D hydrodynamic model for metachronal swimming in which rigid and porous paddles attached to a wall beat asymmetrically, and found an optimal average fluid flux at the same phase lag over an intermediate range of Re $1-800$. It would be useful to study similar models but allowing free swimming (rather than a fixed wall), which includes hydrodynamic interactions generated by the body of the swimmer (not only the paddles). 

In this paper, we computationally study a simple nonreciprocal 2D model swimmer that performs a metachronal stroke. The swimmer is composed of a flat body and two pairs of paddles that beat metachronally with a phase difference $\pi/2$. We investigated the swimmer over a range of paddle separations and Reynolds numbers $Re = 0.05 - 100$ in order to understand the effect of increasing inertia and geometry. We find that for all Re, the swim speed gets smaller when the paddle separation increases and the hydrodynamic interactions between the paddles gets weaker. Interestingly, at all paddle separations, we see a nonmonotonic dependence of the swim speed with respect to Re, indicating the possibility of different motility mechanisms at different Re. Specifically, we see a maximum for $Re\approx1$, a flat minimum between $Re=20-30$ and an eventual increase for $Re>35$. To get more insight, we first study the kinematics of the swimmer, splitting the stroke into power and recovery including for each paddle. The competition between power and recovery strokes of the paddles for parts of the cycle are responsible for the net motion of the swimmer. To show this, we calculate the fluid flows (vorticity and streamlines) averaged over the power stroke, over the recovery stroke and over the full cycle. We identify features in the flow fields that can be linked to the nonmonotonic behavior of the swim speed with $Re$.  

The structure of the paper is as follows. In section II we introduce the swimmer model and describe the methods we use to solve the Navier-Stokes equations numerically and the fluid-structure interactions. In section III, we present our results, first the kinematics, then the fluid flows. We conclude with a discussion in section IV.

\section{Model and Methods}

\begin{figure}
\includegraphics[width=\linewidth]{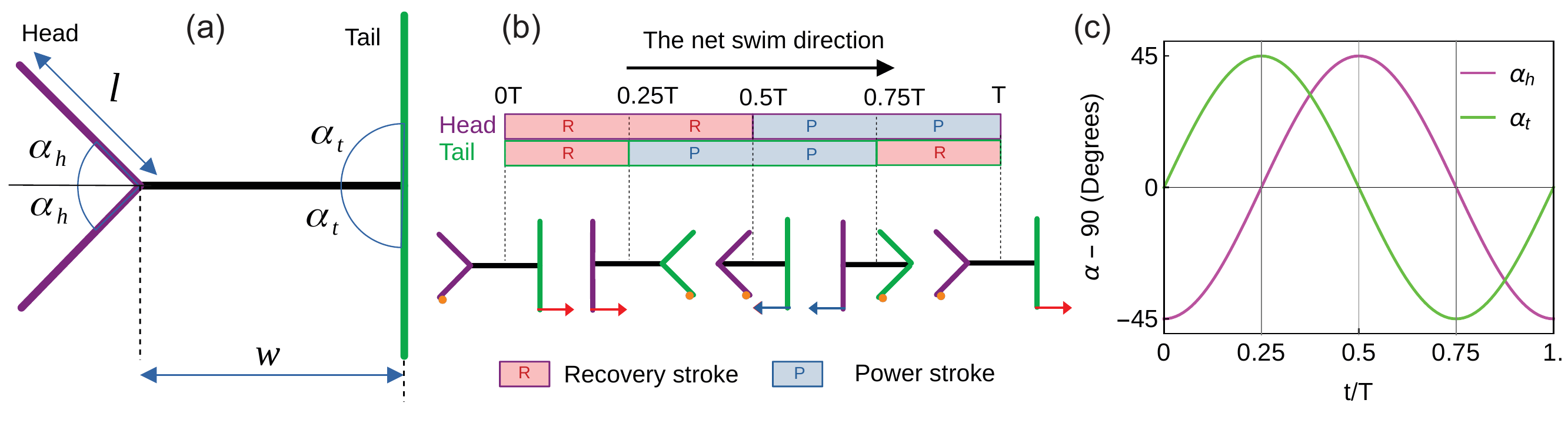}
\caption{(a) A schematic that illustrates the geometry of our model swimmer. All paddles have the same length $l$ and spacing $w$. Angles are specified relative to the body axis which orients horizontally. The dashed vertical line indicates the equilibrium (neutral) angle. (b) Schematic showing the power and recovery strokes of each paddle in one cycle at the beginning of each quarter. Power (recovery) stroke for each paddle is defined in the direction of  decreasing (increasing) angle $\alpha$. Power and recovery stroke are displayed as a blue and red filled box respectively. The red/blue arrows indicate the direction the paddle is moving and the orange dots signify that a paddle is at the end of its oscillation momentarily at rest. (c) Temporal variation of the two angles $\alpha_{h,t}$ relative to the equilibrium (neutral) angle in one complete swim cycle with a phase lag $\phi = \pi/2$ and $A = \pi / 4$.}
\label{fig:model}
\end{figure}

One of the simplest models that can perform a metachronal stroke is composed of a flat plate body and two inflexible pairs of plates for paddles (two on each side), each oscillating around an equilibrium (neutral) angle, see Fig.\ref{fig:model}(a). Note that to isolate the effect of metachronicity and inertia, we employ a symmetric model, where there is temporal (duration of power and recovery stroke), spatial (trajectory of paddles are the same for power and recovery stroke), and orientational symmetry (neutral angle). Thus, the asymmetry that will cause net motion (i.e. swimming) will originate solely from the metachronicity and finite inertia. 
The metachronal beating makes this otherwise symmetric swimmer a nonreciprocal one. 

All paddles have the same length $l$ and the spacing between them is determined by the parameter $w$. Each paddle moves by prescribing the angle $\alpha_{i}$  ($i$ is h:head and t:tail) that it makes with the body according to the following equations:

\begin{equation}
\begin{aligned}
\alpha_h(t) & = \pi/2 + A \cos(2\pi t /T ), \\
\alpha_t(t) & = \pi/2 + A \cos(2\pi t /T + \phi)
,
\end{aligned}
\label{eq:alphaJ}
\end{equation}

\noindent where $A$ and $T$ are the amplitude (in radians) and period of swimming stroke respectively. $\phi$ is the phase lag of the tail paddle relative to the head paddle (therefore, the metachronal wave travels in the tail-to-head direction).

If $\phi = 0$, all paddles beat synchronously, resulting in synchronous swimming. In contrast, metachronal swimming is achieved by setting a non-zero phase lag ($\phi \ne 0$). Previous studies of real organisms have shown that a common phase lag is $\phi= \pi / 2$~\cite{Murphy2011Metachronal} and for a fixed model swimmer, where the body was attached to a physical wall, it was shown that the highest fluid flux is generated with an optimum value of $\phi \approx \pi/2$~\cite{Granzier-nakajima2020ANumerical}.
The same phase lag was used in the robotics model with two paddles operating in the Stokes regime~\cite{ Hayashi2020Metachronal}.
Thus, we will also use  $\phi= \pi / 2$. The prescribed angles $\alpha_{h,t}$ are shown in Fig.~\ref{fig:model}(c).

Each paddle and the swimmer as a whole can be used to define power and recovery strokes. Power and recovery strokes refer to part of the stroke when the limb or swimmer move in the same or opposite direction to swimming. When the limb/swimmer moves in the same direction as swimming, we have the recovery stroke; when the limb/swimmer moves in the opposite direction to swimming, we have the power stroke, see Fig. \ref{fig:model}(b) for a schematic. Because our swimmer is metachronal, power and recovery strokes for the two limbs are not in synch. 

The swimmer is immersed in an initially-stationary Newtonian, incompressible fluid with kinematic viscosity $\nu$ and is assumed to be neutrally buoyant. The Navier-Stokes equations describe the fluid motion: 

\begin{equation}
\begin{aligned}
\rho \frac{D\textbf{u}}{dt} &= -\nabla p + \nu \Delta \textbf{u} + \textbf{f}, \\
\nabla \cdot \textbf{u} &= 0,
\end{aligned}
\label{eq:NSJ}
\end{equation}

\noindent where $D$ is the material derivative, $p$ is the fluid pressure to maintain the incompressibility condition, and $\textbf{u}$ and $\rho$ are the fluid velocity and density respectively. The force due to the presence of the swimmer is $\textbf{f}$; this body force is only applied in the physical region occupied by the swimmer. 

The swimmer is constrained to move along the horizontal axis, which coincides with the body axis; thus, vertical displacement is not allowed. There is symmetry about the swim (or body) axis hence below we show half of the fluid field. We nondimensionalize our variables using the paddle length $l$ as the unit of length and the stroke period $T$ as the unit of time. Initially, we position the swimmer at the center of a simulation domain of size $27 l \times 18l$. 
The size of the simulation box is large enough to eliminate boundary effects, which was checked by performing a series of simulations at different box sizes. The NS equations (\ref{eq:NSJ}) were integrated in time with a time step of $dt = 10^{-4} T$, which  has been chosen to be sufficiently small for computational accuracy and stability.  More details on the computational method and simulation are provided in the Supplemental Information (SI).
A frequency-based Reynolds number (Re) can be defined with respect to the paddle length $l$ and the maximum speed of the tip of the paddles $lA$ in the reference frame fixed to the body (the co-moving frame).
\begin{equation}
Re = \frac{2 \pi A l^2}{\nu T}
\label{eq:Re}
\end{equation}

Equations (\ref{eq:NSJ}) along with the fluid-structure interactions are solved using an efficient and fast algorithm developed by Pantankar and co-workers~\cite{Shirgaonkar2009ANewMathematical} which has been implemented in open-source fluid dynamics platform IBAMR~\cite{Bhalla2013AUnified}. IBAMR is an immersed boundary numerical method with adaptive mesh refinement~\cite{griffith2007adaptive}. The algorithm by Pantankar \textit{et al.} is a variant of a fictitious domain wherein momentum equations for fluid and body are converted to a single equation, which is then solved on a structured Eulerian grid using a fluid solver. It employs distributed Lagrange multipliers to enforce the body deformation kinetics which is given in the co-moving frame. The algorithm then solves for swim velocity, hydrodynamic forces as well as the fluid flow field.

\section{Results}
\subsection{Steady state velocities}

\begin{figure}
\includegraphics[width=5.5 in]{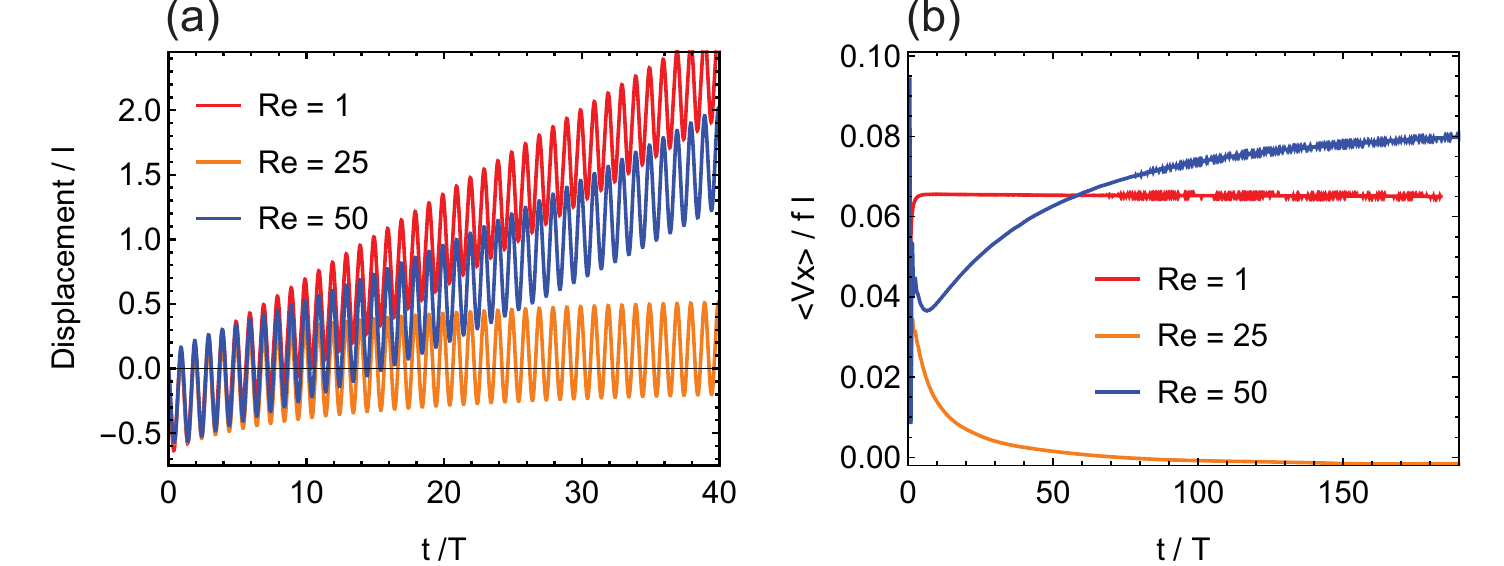}
\caption{Kinematics of swimmer at three characteristic Reynolds numbers, $Re=1,25,50$ for paddle spacing $w = 0.35$ and phase lag $\phi = \pi / 2$. Temporal dependence of (a) the body displacement normalized by the paddle length $l$ and (b) the cycle-averaged velocity of the swimmer normalized by the characteristic speed $l f $, where $f = T^{-1}$ is the frequency, showing steady state.}
\label{fig:KinematicsThreeRe}
\end{figure}

We first present the kinematics for three representative Reynolds numbers, $Re =1, 25, 50$, with fixed paddle spacing $w = 0.35$, phase lag $\phi = \pi / 2$ and beating amplitude $A = \pi / 4$. In all simulations, we fix the paddle length $l$, and the paddle width is ten times smaller.
The swimmer (mid-point of the body) is initially placed at the origin ($x = 0$). Once the simulation starts, all paddles oscillate metachronally resulting in the displacement of the body, see Fig.~\ref{fig:KinematicsThreeRe}(a). At all Re studied here, systems exhibit back-and-forth movement within a single swim cycle, characterizing the jerky motion commonly seen in crustaceans especially in their early life stages~\cite{Williams1994Locomotion, Gauld1959Swimming}. However, the net displacement at the end of each cycle strongly depends on the $Re$ with the largest displacement observed for $Re = 1$ and the smallest for $Re = 25$ (and $Re = 50$ in between the two) indicating nonmotonotic dependence of displacement on Re, see Fig.~\ref{fig:KinematicsThreeRe}(a),(b). Interestingly, in contrast to previous works~\cite{Macmillan1981Coordination, Takagi2015Swimming, Kohlhage1994AnAnalysis}, the direction of net motion is opposite to the metachronal wave, which travels from tail-to-head or (the negative x), so our model swims in the direction of its tail.

We monitor the velocity of the swimmer at different $Re$ and we see that after an initial transient time when the flow is still developing, our systems reach steady state velocities $\left< V_x \right>$, in which variation of the cycle-averaged velocity with time is small, less than $2\%$ over consecutive cycles, see Fig. \ref{fig:KinematicsThreeRe}(b). The time it takes to reach steady state increases with increasing $Re$; at $Re = 1$ it takes about $O(10)$ oscillations while it takes $O(100)$ at higher $Re$. As a reference, the steady-state speeds for $Re = 1, 50$ are $0.065$ and $0.083 $ respectively. Interestingly, the $Re = 25$ swimmer eventually exhibits no net motion. To get insight into the swimmer's locomotion, it is informative to look at the instantaneous displacement and velocity throughout one complete cycle, after steady state has been reached.

\begin{figure}
\includegraphics[width=3.5 in]{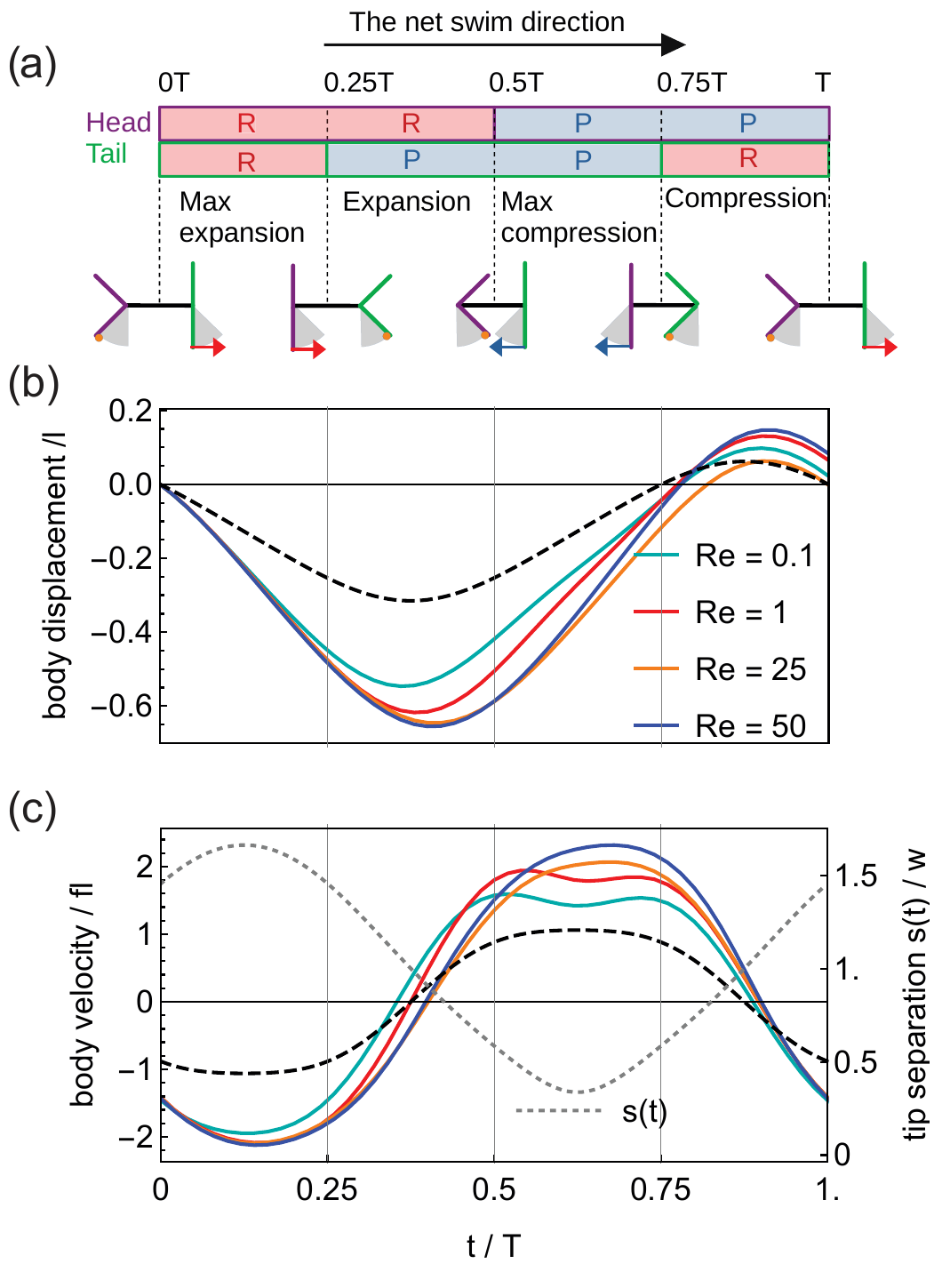}
\caption{Swimmer kinematics during one cycle at four representative $Re=0.1, 1, 25, 50$ for paddle spacing $w = 0.35$ and phase lag $\phi = \pi / 2$. (a) Schematic showing power and recovery strokes for each paddle. P and R stand for power and recovery respectively. The grey shaded areas indicate areas swept by individual paddles during that quarter. 
(b) The normalized body displacement over a cycle. 
(c, left axis) Normalized instantaneous body velocity $v_b(t)$ along the swim axis. The dotted curve in (c, right axis) shows the paddle tip separation $s(t)$ normalized by paddle spacing $w$ over a cycle. 
The dashed black curves in (b) and (c) are the displacement and velocity of the swimmer in the co-moving frame.
}
\label{fig:KinematicsCompleteCycle}
\end{figure}

The instantaneous displacement and velocity of the swimmer over one cycle of oscillation for four characteristic $Re=0.1,1,25,50$ are shown in Fig.\ref{fig:KinematicsCompleteCycle}(b-c). 
Since our model swimmer is  force- and torque-free and self-propels by moving its paddles (i.e. paddle kinematics is prescribed in the reference frame fixed to the body or co-moving frame), we can write down an expression for body displacement and velocity in the co-moving frame using conservation of momentum. Given the variation of angle $\alpha$ in eq. (\ref{eq:alphaJ}) and assuming a uniform mass distribution,  the body's velocity is: 

\begin{equation}
v^0_{b}(t) = - \frac{l^2}{4l + w} \left[
\frac{\partial \alpha_h}{\partial{t}} \sin (\alpha_h) +
\frac{\partial \alpha_t}{\partial{t}} \sin (\alpha_t)
\right]
\label{eq:bodyvel}
\end{equation}
 
\noindent and the associated displacement $d^0_{b}(t) = \int_{0}^{t} v^0_{b}(t^{\prime}) d t^{\prime}$. Note that the prefactor $(4 l + w)$ in eq. (\ref{eq:bodyvel}) is just the total body length of swimmer. The superscript `$0$' denotes quantities that are expressed in the co-moving frame.
In general, they only coincide with their counterparts in the lab frame in the limit of no fluid present, thus they are Re-independent. $v^0_{b}(t)$ and $d^0_{b}(t)$ are plotted as dashed curves in Fig.~\ref{fig:KinematicsCompleteCycle}(b-c).
We also plot the separation of the tips of the paddles, in order to see how that relates to the kinematics of the swimmer, see dotted line Fig.~\ref{fig:KinematicsCompleteCycle}(c).

In the first quarter $t = 0 - 0.25 T$, both head and tail paddles are in the recovery stroke, they collectively push fluid to the right, thus the body translates to the left (negative x) leading to the body's negative displacement, see Fig.\ref{fig:KinematicsCompleteCycle}(b). This is the quarter where the paddles have the widest separation, i.e. we see maximum expansion. Displacement (and velocity) are insensitive to Re except for a slight dependence at low $Re = 0.1$. The velocity of the swimmer reaches a minimum in the negative direction.

In the second quarter $t = 0.25 - 0.5 T$, the head paddle is still in recovery stroke while the tail is now in power stroke; so head and tail move in opposite directions and thus have opposing effects on the net motion of the swimmer. The swimmer slows down and switches direction in the middle of the quarter to swim forward, and the backward displacement reaches a minimum. In this quarter we start to see distinction between different Reynolds numbers for both displacement and velocity. The backward displacement is the largest for the higher Reynolds numbers (Re$=25, 50$) and smallest for Re$=0.1$. This is surprising and different from reciprocal models. For example, the backward displacement for a dumbbell swimmer~\cite{Dombrowski2019Transition,Dombrowski2020Kinematics} is larger the lower the Re, and is largest for Stokes flow. This makes sense, because the recovery stroke for a reciprocal swimmer is identical and opposite to the power stroke so the swimmer goes back-and-forth more, when the inertia is less (ultimately turning into Purcell's scallop with no net motion in Stokes flow~\cite{Purcell1977}). As Re increases both curves shift slightly to the right, to later times -- the delay can be interpreted as the effect of increasing inertia.

The next quarter $t = 0.5 - 0.75 T$, occurs with both paddles in power stroke. The paddles  push fluid to the left, thus the swimmer reaches maximum speed in the direction of swimming to the right. Paddle tip separation reaches a minimum here, i.e. we see maximum compression. All velocity curves reach a maximum in the swim direction. We speculate that the double peak at low Re is related to the instantaneous fluid flows, see SI. It is interesting that here, we see a big difference amongst velocity curves at different Re, showing how much inertia plays a role. We also see a difference between the velocity curves with respect to the reference case without fluid, suggesting a strong effect of hydrodynamic interactions between paddles.

In the last quarter $t = 0.75 - 1T$, the two paddles oscillate in opposite directions again, the head performing a power stroke, the tail a recovery stroke. For each Re, we see a different displacement, which is also the net displacement over the whole cycle. The swimmers with the largest net displacements are Re$=50$ and Re$=1$, the ones with the lowest are $Re=0.1$ and $Re=25$. All curves reach the final displacement with the same velocity at $t = T$.  

Taken together, Fig. \ref{fig:KinematicsCompleteCycle} indicates that the velocity fluctuations over a cycle increase with increasing inertia (i.e. Re). This means more inertial energy spent~\cite{Kwak2017Design}. Additionally, the backward motion is present for all $Re$, and it increases as $Re$ increases, as shown in Fig.~\ref{fig:KinematicsCompleteCycle}(b,c). This is surprising when compared to reciprocal swimmers where the backward motion monotonically gets smaller as Re increases. One of the reasons is probably because the swim stroke of each paddle individually is symmetric; power and recovery stroke for each paddle are mirrored to one other with equal time duration spent for each.
A previous study~\cite{Kwak2017Design} of a similar metachronal swimming model has shown that introduction of asymmetry between the power and recovery stroke of individual paddles (e.g. bending during recovery) can reduce or even eliminate backward translation. However, as noted in that paper, reduction of backward movement does not generally lead to an increase in steady state swim speed. 
A closer examination of the $Re = 25$ curves reveals that the velocity variation is nearly symmetric across the time axis. As a result, the net displacement for $Re = 25$ is vanishingly small as observed in Fig. \ref{fig:KinematicsCompleteCycle}(b).

\begin{figure}
\includegraphics[width=4 in]{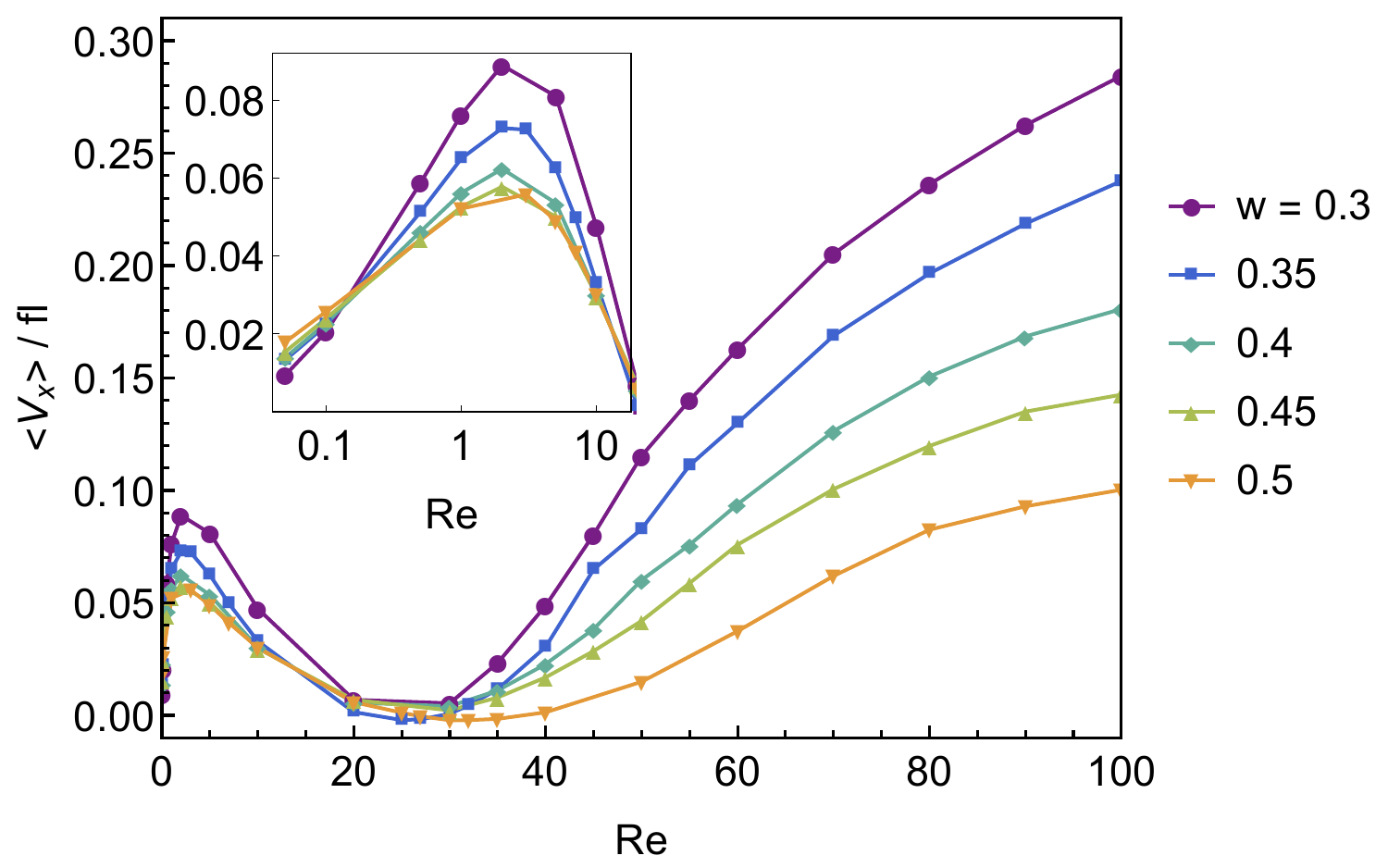}
\caption{Nonmonotonic variation of the swim speed $\langle V_x \rangle$ normalized by the characteristic speed ($f l$, where $f$ is the frequency) of the swimmer as a function of $Re$ for various paddle spacings $w$. $\langle V_x \rangle$ is calculated as the net displacement over one cycle at steady state. The inset is a zoom-in at small $Re$ plotted on a semilogarithmic scale.}
\label{fig:avgVvsRe}
\end{figure}

\subsection{Effect of inertia and geometry on the steady state velocities}
To get more insight on the effect of increasing inertia on our swimmer, we perform simulations for a range of Reynolds numbers between $0.05\le Re \le100$ and we also vary the paddle spacing $w$, see Fig. \ref{fig:avgVvsRe}. The swim velocity of the swimmer shows an interesting nonmonotonic dependence on Re. At $Re \gtrsim 0$ close to the Stokes regime, the swimmer's cycle-averaged velocity $\langle V_x \rangle$ is negligibly small, consistent with recent experiments on a robotics model~\cite{Hayashi2020Metachronal}.
As inertia is introduced ($Re > 0$), $\langle V_x \rangle$ increases sharply until $Re \approx 1-3$, where all systems exhibit a peak. After that, $\langle V_x \rangle$ gradually drops to a flat dip between $Re\approx 20 - 30$ before it increases rapidly with increasing $Re$, at a rate inversely proportional to the paddle spacing $w$. The dip width between $Re\approx 20-30$ flattens as the paddle spacing increases from $w = 0.3$ to $w = 0.5$. For all $Re>0.1$ the speed of the swimmer drops with increasing paddle spacing $w$. In other words, the dependence of $\langle V_x \rangle$ on $Re$ is stronger for smaller spacing. This result is consistent with the expectation that a stronger hydrodynamic interaction should occur between tight-spacing paddles.
The nonmonotonic behavior of our swimmer's velocity with $Re$ suggests that there may be different mechanisms at different $Re$, something that seems to be a signature of intermediate-Re dynamics, see for example a transition in the swim direction for a reciprocal swimmer at a similar $Re \approx 20$. To understand better the motility mechanism(s) here, we next consider the associated fluid flows.

\begin{figure}
\includegraphics[width=\textwidth]{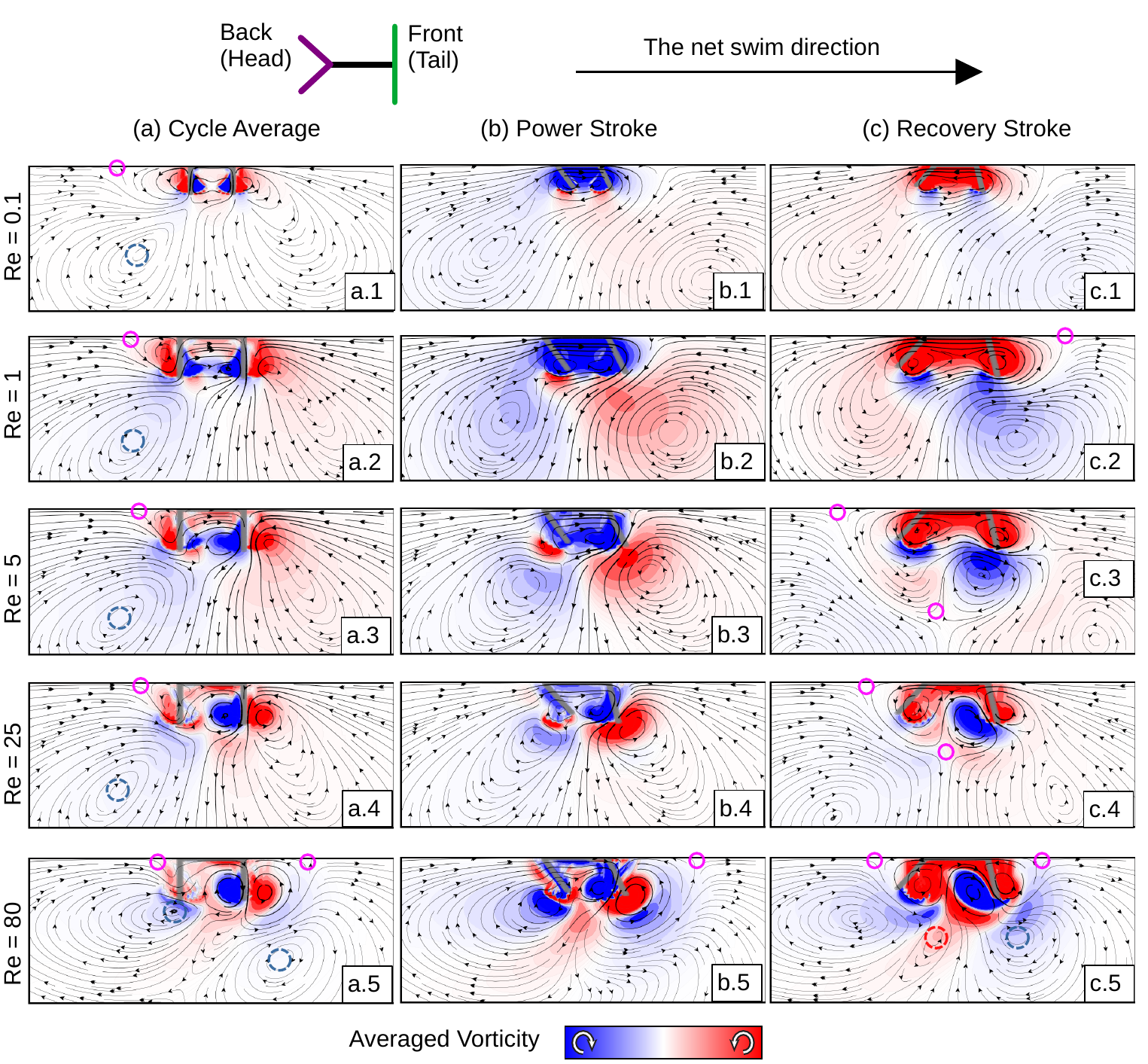}
\caption{Streamline (black arrows) and vorticity (color map) fields are plotted, averaged over (a) a complete cycle, (b) the power stroke and (c) the recovery stroke of the swimmer, at characteristic $Re=0.1, 1, 5, 25, 80$ with paddle spacing $w = 0.35$. 
The location of the swimmer is shown in gray; at the neutral position in (a) and at the start of each interval, power-stroke interval in (b), recovery-stroke interval in (c). 
Due to symmetry, only bottom half of the swimmer and flow field are shown. 
The dashed open circles indicate the location of key vortices: blue and red for clockwise and counterclockwise rotating vortices respectively. The magenta open circles indicate specific stagnation points. The averaged vorticity is normalized by its maximum for the positive and minimum values for the negative.}
\label{fig:avgFlow}
\end{figure}

\subsection{Flow field in different swimming regions.}

Since the cycle-averaged speed is just the resultant effect of the back-and-forth motion, it is natural to decompose each swim cycle into a power and recovery stroke for the entire swimmer as well (similar to the ones defined for individual paddles). We define power when the swimmer's velocity is in the direction of swimming and recovery when the swimmer's velocity is in the opposite direction to swimming, positive and negative respectively in Fig.3(c). Note that when considering the whole swimmer, the power and recovery strokes do not necessarily have the same time duration and they can vary with $Re$. Nonetheless, in Fig.3(c), we can see that the recovery strokes for the swimmer occur as the paddles expand, and the power strokes occur as the paddles compress, see paddle tip separation. 

We calculate the streamline and vorticity of the fluid flow averaged over the entire cycle, as well as averaged over power and recovery strokes separately for four characteristic Reynolds numbers, $Re=0.1, 1, 5, 25, 80$, see Fig.~\ref{fig:avgFlow}. Considering the cycle-averaged flow fields first, we see that the far field for all but the highest Re, resembles that of a puller, where fluid is pulling in along the swim axis and pushing out in perpendicular, see Fig.~\ref{fig:avgFlow}(a.1-4). At all Re, we see a front-back asymmetry in the flow fields, which is expected, given that the swimmer is nonreciprocal. For example, for Re=0.1 (Fig.~\ref{fig:avgFlow}(a.1)), there are two large counter clockwise (CCW) vortices one on the front, i.e. right hand side (RHS), and one in the back, i.e. left hand side (LHS) of the swimmer, but the one at the back is spatially smaller and meets with a clockwise (CW) vortex -- the point where they meet creates a stagnation point along the swim axis (indicated by the solid circle in Fig.~\ref{fig:avgFlow}(a.1)). It is further than that stagnation point that the $Re=0.1$ field can be considered to be puller-like. As Re increases the boundary layer thickness gets smaller and we expect to see vortices confined closer to the surface of the swimmer. Indeed, the stagnation point, that signifies the size of the CCW vortex on the LHS of the swimmer in (a.1), moves closer and closer to the swimmer surface, see Fig.~\ref{fig:avgFlow}(a.1-5). In front of the swimmer, we see something similar: the CCW vortex moves closer to the swimmer surface and becomes more spatially confined until at $Re=80$ we see the appearance of a CW vortex and a new stagnation
point, Fig.~\ref{fig:avgFlow}(a.5). The nonmonotonic behavior of the velocity as a function of Re in Fig.4 could indicate that there is a qualitative difference between swimming at $Re\lessapprox25$ versus $Re\gtrapprox25$. So, even though the changes we see here in the flow fields are gradual, there is a qualitative distinction between $Re=0.1, 1, 5$ versus $Re=25, 80$, where for the latter two, we see the development of two counter-rotating vortices on either side of the tail paddle (that are not present for the head paddle).

Next, we consider the flow fields for each Re, analyzing not only the cycle-averaged flow but also the flows averaged over the power and recovery strokes separately. 
At $Re = 0.1$ (Fig. \ref{fig:avgFlow}(a-c.1), this is a mostly viscous-dominated region, where vortices created are quickly dissipated into the fluid, hence the flow fields between power and recovery seem close to equal and opposite even though they are not, (Fig. \ref{fig:avgFlow}(b-c.1)). Two CW vortices are generated between the paddles, which give rise to a CW circulation of fluid inside the paddle cavity, see Fig.~\ref{fig:avgFlow}(a.1). The stagnation point on the LHS of the swimmer in (a.1) seems to occur because of the reversal of the flow in that region between power and recovery strokes. During the power stroke, i.e. compression of the paddles, the flow is pushed outward from the paddle cavity (roughly perpendicular to the swimmer body) and inward along the swim axis. During the recovery stroke, i.e. expansion of the paddles, the flow is moving in toward the cavity on the perpendicular to the swim axis and pushing out along it.    

The averaged flow fields for the power stroke of $Re = 1$ and $5$ look qualitatively similar, but their counterparts for their recovery stroke are quite different, see Fig. \ref{fig:avgFlow}(b.2-3),(c.2-3). Consider the recovery strokes: the CW vortex below the tail paddle shrinks as inertia increases from $Re = 1$ to $5$ (Fig. \ref{fig:avgFlow}(c.2-3)); consequently, the stagnation point on the RHS of the tail paddle at $Re = 1$ (Fig. \ref{fig:avgFlow}(c.2)), disappears at $Re = 5$ (Fig. \ref{fig:avgFlow}(c.3)). As a result, the flow on the RHS along the swim axis is reversed from pushing away to the right to pulling inward to the left. Thus, it seems that there is a lot more fluid in the backward direction towards the left during the recovery stroke at $Re=5$ than at $Re=1$, (Fig. \ref{fig:avgFlow}(c.3-4)). The net result is that the cycle-averaged speed of the swimmer is reduced as $Re$ increases from $1$ to $5$. This mechanism is persistent at even higher $Re$, possibly explaining why the swimmer speed drops in the flat minimum between $Re = 1 - 30$ observed in Fig.~\ref{fig:avgVvsRe}. 
Another example of a shrinking vortex is the CCW one on the RHS and below the swimmer during the power stroke, between $Re=0.1$ and $Re=80$, see Fig.~\ref{fig:avgVvsRe}(b.1-5).

It is interesting to make a distinction in the features of the fluid fields that are oscillatory versus features that are present at all times during the cycle. The oscillatory features switch back and forth between power and recovery strokes and are more evident at low Re, \textit{e.g.} the fields of power and recovery for $Re=0.1$ look qualitatively almost equal and opposite to one another, see Fig.~\ref{fig:avgVvsRe}(b.1)(c.1). But as Re increases, we see the development of vortices that are present at all times. For example, the CCW vortex on the RHS of the swimmer, for $Re=5$ and $Re=25$ is present during power, recovery and the cycle-averaged, see Fig.~\ref{fig:avgVvsRe}(a-c.3),(a-c.4). Note that we can confirm this by looking at  instantaneous flows shown in the SI. Another example are the CW vortices below the swimmer at $Re=80$, Fig.~\ref{fig:avgVvsRe}(a-c.5). These nonoscillatory steady features are reminiscent of steady streaming flows~\cite{riley66,riley2001}, which have also been studied in the context of motility~\cite{Dombrowski2020Kinematics}.  
In general, we notice, that though the flows seem to reverse between power and recovery, it is the power stroke that resembles more closely the cycle-averaged flow field, \textit{i.e.} the field that dominates. 

We note that if paddle-paddle or paddle-body hydrodynamic interactions are ignored in our model, then each paddle will develop its own local symmetric flow, yielding no net motion as analytically derived before ~\cite{Takagi2015Swimming}. 
However, since hydrodynamic body-fluid interaction is fully accounted for in our model, it is not surprising that swimmer velocity is non-zero (but very small) even at the lowest $Re$ tested, $Re = 0.05$

\section{Discussion}

To sum up, we computationally studied a simple nonreciprocal metachronal swimmer at intermediate Reynolds numbers, inspired by brine shrimp. We found, that even for such a simple model, consisting of a flat body and two pairs of paddles, there is interesting and surprising behavior. Specifically, we found that the swim speed depends nonmonotonically on Re, indicating different regimes of motility mechanisms. We were able to relate this behavior to the fluid flows averaged over a cycle and separately averaged over the power and recovery strokes. 

Nonmonotonic behavior of the swim speed with respect to Re seems to be a signature of the intermediate Reynolds range. In our previous work, we found a switch in the swim direction for a simple reciprocal dumbbell swimmer~\cite{Dombrowski2019Transition,Dombrowski2020Kinematics} -- we do not see a switch here, but we do see what seems to be different subregimes. It is also interesting that although the two models are geometrically very different (asymmetric dumbbell in ~\cite{Dombrowski2019Transition,Dombrowski2020Kinematics} versus a flat plate with paddles here), the swim speeds as a function of Re look remarkably similar: they both have a maximum at low $Re\approx1$ and the speed vanishes around $Re\approx20$ to increase again as $Re$ increases further (see Fig.2 in ~\cite{Dombrowski2019Transition}). In both cases, it is the frequency Reynolds number that is being referred to.   

We emphasize the importance of developing and studying simple models to understand motility at intermediate Re in order to formulate unifying principles, trends and behaviors. In future investigations, it would be interesting to add complexity to the model presented here, to resemble the brine shrimp organism more, \textit{e.g.} by adding more paddles (legs), or allowing bending of paddles. A question that arises also is whether real swimmers display this kind of nonmonotonic behavior and, if not, how do they adjust their swimming to avoid it. Finally, interactions between multiple swimmers at intermediate Reynolds numbers is a yet-unexplored field, where models like ours presented here could provide many insights.

\acknowledgments
D.K. and H.N. acknowledge the National Science Foundation, grant award DMR-1753148. 

\bibliography{brineshrimp}
\end{document}